\documentstyle[12pt]{article}
\newtheorem{theorem}{Theorem}
\newtheorem{itlemma}{Lemma}[section]
\newtheorem{itproposition}[itlemma]{Proposition}
\newtheorem{itcorollary}[itlemma]{Corollary}
\newtheorem{itremark}[itlemma]{Remark}
\newtheorem{itremarks}[itlemma]{Remarks}
\newtheorem{itdefinition}[itlemma]{Definition}
\newtheorem{itexample}[itlemma]{Example}

\newenvironment{lemma}{\begin{itlemma}\rm}{\end{itlemma}} 
\newenvironment{remark}{\begin{itremark}\rm}{\end{itremark}} 
\newenvironment{remarks}{\begin{itremarks} \rm}{\end{itremarks}}
\newenvironment{corollary}{\begin{itcorollary}\rm}{\end{itcorollary}}
\newenvironment{proposition}{\begin{itproposition}\rm}{\end{itproposition}}
\newenvironment{definition}{\begin{itdefinition}\rm}{\end{itdefinition}}
\newenvironment{example}{\begin{itexample}\rm}{\end{itexample}}
\newenvironment{fact}{\noindent {\em Fact}. \ \ }{\hfill \medskip}
\newenvironment{proof}{\noindent {\em Proof}.\ \
}{\hspace*{\fill}$\Box$\medskip}
\newenvironment{claim}{\noindent {\em Claim}. \ \ }{\hfill \medskip}
\newcommand{\be}[1]{\begin{equation}\label{#1}}
\newcommand{\ee}{\end{equation}}
\newcommand{\bl}[1]{\begin{lemma}\label{#1}}
\newcommand{\br}[1]{\begin{remark}\label{#1}}
\newcommand{\brs}[1]{\begin{remarks}\label{#1}}
\newcommand{\bt}[1]{\begin{theorem}\label{#1}}
\newcommand{\bd}[1]{\begin{definition}\label{#1}}
\newcommand{\bp}[1]{\begin{proposition}\label{#1}}
\newcommand{\bc}[1]{\begin{corollary}\label{#1}}
\newcommand{\bfact}[1]{\begin{fact}\label{#1}}
\newcommand{\bex}[1]{\begin{example}\label{#1}}
\newcommand{\ec}{\end{corollary}}
\newcommand{\efact}{\end{fact}}
\newcommand{\eex}{\end{example}}
\newcommand{\el}{\end{lemma}}
\newcommand{\er}{\end{remark}}
\newcommand{\ers}{\end{remarks}}
\newcommand{\et}{\end{theorem}}
\newcommand{\ed}{\end{definition}}
\newcommand{\ep}{\end{proposition}}
\newcommand{\epr}{\end{proof}}
\newcommand{\bpr}{\begin{proof}}
\newcommand{\bcl}{\begin{claim}}
\newcommand{\ecl}{\end{claim}}

\newcommand{\bi}{\begin{itemize}}
\newcommand{\ei}{\end{itemize}}
\newcommand{\ben}{\begin{enumerate}}
\newcommand{\een}{\end{enumerate}}
\newcommand{\text}[1]{\hbox{\rm \ #1\ \/}}

\newcommand{\CC}{\mbox{${\rm \:  C\!\!\! I
\;\;}$}}

\newcommand{\RR}{\mbox{${\rm \:  R\!\!\!\! I
\;\;}$}}

\newcommand{\vs}{\vspace{0.25cm}}
\newcommand{\qed}{\hfill $\Box$ \vskip 2ex}

\begin{document}
\noindent \hfill {\large{}}
\begin{center}
{\Large NOTIONS OF CONTROLLABILITY
FOR QUANTUM MECHANICAL SYSTEMS}
\end{center}

\bigskip

\begin{center}

{Francesca Albertini\\
\vs
Dipartimento di Matematica
Pura ed Applicata,\\
Universita' di Padova,\\
via Belzoni 7,\\
35100 Padova, Italy.\\
Tel. (+39) 049 827 5966\\
email: albertin@math.unipd.it}\\

\vs
\vs
{Domenico D'Alessandro \\
\vs
Department of Mathematics\\
Iowa State University \\
Ames, IA 50011,  USA\\
Tel. (+1) 515 294 8130\\
email: daless@iastate.edu}

\end{center}

\vspace{0.5cm}

\begin{abstract}

In this paper,  we define four  different notions of controllability of
physical interest for
 multilevel quantum mechanical systems. These notions
involve the possibility of driving the evolution operator as well
as the state of the system. We establish the connections among
these different notions as well as
 methods to
verify controllability.

The paper also contains results on the relation between the
controllability in arbitrary small time of a system varying on a
compact transformation Lie group and the corresponding system on
the associated homogeneous space
with applications to quantum systems.

\end{abstract}

\section{Introduction}
In this paper, we consider multilevel quantum system described by
a finite dimensional bilinear model  \cite{conmoh} \cite{murti}
\be{gensys} \dot
{|\psi>}=(A+\sum_{i=1}^mB_iu_i(t)){|\psi>}, \ee where
$|\psi>$\footnote{In this paper, we use Dirac notation  $|\psi>$
to denote a vector on $\CC^n$ of length $1$, and
$<\psi|:=|\psi>^*$ where $^*$ denotes transposed conjugate} is the
state  vector varying on the  complex sphere $S^{n-1}_{\CC}$,
defined as the set of $n$-ples of complex numbers $x_j+iy_j$,
$j=1,...,n$, with $\sum_{j=1}^n x_j^2+y_j^2 =1$. The matrices $A,$
$B_1,...,B_m$ are in the Lie algebra of {\it skew-Hermitian}
matrices of dimension $n$, $u(n)$. If $A$ and $B_i$, $i=1,...,m$
have zero trace they are in the Lie algebra of skew Hermitian
matrices with zero trace, $su(n)$. The functions $u_i(t)$,
$i=1,2,...,m$ are the controls. They are  assumed to be piecewise
continuous and unconstrained in magnitude, although this
assumption is immaterial for most of the theory developed here. Models
of quantum control systems different from the
bilinear one (\ref{gensys}) may be more appropriate in some cases
(see e.g. \cite{DKAB1}, \cite{DKAB2}). The
controllability for infinite
dimensional bilinear systems
has been studied in \cite{Ball} \cite{Tarn}.

The solution of (\ref{gensys}) at time
$t$, $|\psi(t)>$ with initial condition $|\psi_0>$, is given by:
\be{solutione}
|\psi(t)>=X(t)|\psi_0>,
\ee
where $X(t)$ is the solution at time $t$ of the equation
\be{jasbd}
\dot X(t)=(A+\sum_{i=1}^mB_iu_i(t))X(t),
\ee
with initial condition $X(0)=I_{n \times n}$. The matrix
$X(t)$ varies on the Lie group
of special unitary matrices $SU(n)$ or the Lie group
of unitary matrices $U(n)$ according
to whether or not the matrices $A$ and $B_i$ in
(\ref{jasbd}) have all zero trace.

\vs

The controllability of the system (\ref{gensys}) is usually
investigated by applying general results on bilinear right
invariant systems on compact Lie groups \cite{SUSS} \cite{murti}.
These results, applied to our model,  give a necessary and
sufficient condition for the set of states reachable for system
(\ref{jasbd}) to be the whole Lie group $U(n)$ (or $SU(n)$). The
condition is given in terms of the Lie algebra generated by the
matrices $A$, $B_1,$...,$B_m$. Since both the groups $U(n)$ and
$SU(n)$ are transitive on the complex sphere, it follows {}from
(\ref{solutione}) that  this condition is also a sufficient
condition for the controllability of  the state
$|\psi>$. Controllability results for quantum systems that do not use
the Lie algebraic approach have been developed in \cite{Kime},
\cite{Turinici1} \cite{Turinici2}. Investigations related to the one
presented here were carried out in \cite{Schirmer1} \cite{Schirmer2}
\cite{Schirmer3} \cite{Schirmer4} which also present a number of
examples of applications.

\vs

In this paper, we first define four different notions of
controllability which are of physical interest for quantum
mechanical systems of the form (\ref{gensys}).  Using general
results on transitivity of transformation groups, we provide
criteria to check these  controllability notions and we establish
the connections among them. This is done in Sections 2 through 7.
Then, we investigate the relation between various notions of
controllability in arbitrary small time and prove some  general results
in Section 8, which relates controllability in arbitrary time for
a system varying on a compact transformation Lie group and the
corresponding system on the associated homogeneous space. Although
this
 is motivated by the analysis of multilevel quantum systems
(where the transformation group is $SU(n)$ and the homogeneous
space is the sphere $S_{\CC}^{n-1}$) the analysis presented in
Section 8 is valid for any compact transformation Lie group. As an
application, we consider the important model of two
interacting spin $\frac{1}{2}$ particles
in an electro-magnetic field for which we
show that not every state transfer
can be obtained
in arbitrary small
time. Conclusions
are given in
Section 9.

\section{Definitions of Notions of Controllability
for Multilevel Quantum Systems}

The following notions of controllability
are of physical interest for
quantum mechanical systems
described in (\ref{gensys}):

\begin{itemize}

\vs

\item {\bf Operator-Controllability (OC)}. The system
is {\it {operator-controllable}} if every desired
 unitary (or special unitary) operation on the
state can be performed using an
appropriate control field. {}From (\ref{solutione}) and
(\ref{jasbd}), this means that there exists an admissible control
to drive the state $X$ in (\ref{jasbd}) {}{}from the Identity to
$X_f$, for any $X_f \in U(n)$ (or $SU(n)$).

\end{itemize}

\noindent We shall use the term
operator controllable for both the unitary case and the special unitary
case pointing out the difference between  the two cases wherever
appropriate. Operator controllability in the unitary case is called
`Complete Controllability' in \cite{Schirmer1}-\cite{Schirmer2}

\begin{itemize}

\item{\bf Pure-State-Controllability (PSC)} The system
is {\it{pure-state-controllable}} if
 for every pair of initial
and final states, $|\psi_0>$ and
$|\psi_1>$ in $S^{n-1}_{\CC}$ there exist  control functions $u_1,...,u_m$
and a time $t>0$ such that the solution of (\ref{gensys}) at time $t$,
with initial
condition $|\psi_0>$, is  $|\psi(t)>=|\psi_1>$.

\vs

\item{\bf Equivalent-State-Controllability (ESC)} The system is
{\it{ equivalent-state-control-\linebreak lable}} if, for every pair
 of initial and final states,
$|\psi_0>$ and $|\psi_1>$ in $S^{n-1}_{\CC}$,  there exist
controls $u_1,...,u_m$ and a phase factor $\phi$ such that the
solution of (\ref{gensys}) $|\psi>$, with $|\psi(0)>=|\psi_0>$,
satisfies
$|\psi(t)>= e^{i\phi}|\psi_1>$, at some $t >0$.
\end{itemize}
A density matrix $\rho$ is a matrix of the form
$\rho:=\sum_{j=1}^r w_j |\psi_j><\psi_j|$, where the coefficients
$w_j>0$, $j=1,2,...,r$, satisfy $\sum_{j=1}^r w_j=1$ (see e.g.
\cite{sakurai} Chp. 3). The state of a quantum system can be described
by a density matrix. In particular, this is necessary
when the system is an ensemble of a number of non
interacting quantum systems. The constants $w_{j}$, $j=1,\ldots,r$,
give the proportion of such systems in the pure states $|\psi_{j}>$.
\begin{itemize}

%
\item{\bf Density-Matrix-Controllability (DMC)} The system
is density matrix controllable if, for each pair of unitarily
equivalent \footnote{Two matrices $A,\, B\in U(n)$ are said to be unitarily
equivalent if there exists a matrix $C\in U(n)$ such that
$CAC^{*}=B$} density matrices $\rho_1$ and $\rho_2$, there exists a
control $u_1,u_2,...,u_m$ and a time $t>0$, such that the solution of
(\ref{jasbd}) at time $t$, $X(t)$,  satisfies
\be{DMC}
X(t)\rho_1 X^*(t)=\rho_2.
\ee

\end{itemize}

Equivalent state controllability is of interest because, in quantum
mechanics, states that differ by  a phase factor are physically
indistinguishable. Therefore, {}from a physics point of view,
 having $ESC$ is as good as having $PSC$. Density matrix controllability
is of interest when a mixed ensemble of different states is
considered. In this case, the state at every
time is represented by a
density matrix which evolves as $\rho(t)=X(t) \rho(0) X^*(t)$, where
$X(t)$ is solution of (\ref{jasbd}) with initial
condition equal to the identity.  Since
$X(t)$ is unitary, only density matrices that are unitarily
equivalent to the initial one can be obtained through time evolution.

\vs

In the following five   sections we
study the previous four notions of
controllability, give criteria
to check them in practice,
and discuss the relation among them.

\section{Operator Controllability}

Operator controllability is the type of controllability considered
in \cite{murti}.   Operator controllability can be checked by
verifying the {\it Lie algebra rank condition} \cite{SUSS}, namely
by verifying whether or not the Lie algebra generated by
$\{A,B_1,B_2,...,B_m\}$ is the whole Lie algebra $u(n)$ (or
$su(n)$).  More in general, recall that there exists a one to one
correspondence between the Lie subalgebras of $u(n)$ and the
connected Lie subgroups of $U(n)$. We will denote in the sequel
by $\cal L$ the Lie
algebra generated by
 $\{ A, B_1,B_2,...,B_m \}$ and by $e^{\cal L}$ the corresponding connected
Lie
subgroup of $U(n)$. We have the
following result, which, in essence, follows {}from the
fact that $U(n)$ is a {\it compact} Lie group.

\vs

\bt{1} The set of states attainable {}from the Identity for
system (\ref{jasbd}) is given by the connected Lie subgroup
$e^{\cal L}$, corresponding to the Lie algebra ${\cal L}$,
generated by $\{A,B_1,B_2,...,B_m\}$.
\et
\bpr
It is clear that we can look at the system
(\ref{jasbd}) as having state
 varying on $e^{\cal L}$. The
topology on $e^{\cal L}$ is the one induced by the one of $U(n)$.
First, we show that the identity $I$ is a Poisson stable point
(see \cite{Kunita}) for the flow $e^{At}$ on $e^{\cal L}$. Assume,
by the way of contradiction, that $I$ is not a Poisson stable
point. Then there exists a time $T$ and an open set
$K:=B(I,\epsilon) \cap e^{\cal L}$, with $B(I,\epsilon)$ a ball in
$U(n)$ of radius $\epsilon$, such that $e^{At}$  is not an element
of $K$ for any $t>T$. In particular, for $t>T$, $e^{At}$ is never
an element of $B(I, \epsilon)$.  Fix $t>T$ and consider the ball
$H=B(I,\frac{\epsilon}{2})$ in $U(n)$. We have, for any $n
\not=j$, $n,j$ positive integers, $$ e^{nAt} H \cap e^{jAt}H
=\emptyset. $$ If this was not the case, we would have had that
the distance between $e^{kAt}$, $k:=|n-j|$, and $I$ would have
been less than $\epsilon$ contradicting what we have said before.
Therefore we have found an infinite sequence of disjoint open
balls of the same radius which contradicts compactness of the
whole space $U(n)$. Thus $I$ is a Poisson stable point. Using this
fact, we may apply Theorem 4.4 of \cite{Kunita} to conclude that
the set attainable {}from the identity for system (\ref{jasbd}) is
$e^{\cal L}$. \epr

{}From Theorem \ref{1}, it is clear that the Lie algebra rank
condition is also necessary to have operator controllability,
thus we have: \bc{coro1} System (\ref{jasbd}) is
operator-controllable if and only if ${\cal{L}}= u(n)$ (or
${\cal{L}}= su(n)$). \ec

\section{Pure State Controllability}

{}From the representation of the solution of Schr\"odinger equation
(\ref{solutione}), it is clear that the system is
pure state controllable if and only if the Lie group $e^{\cal L}$
corresponding to the Lie algebra ${\cal L}$ generated by
$\{A,B_1,...,B_m\}$ is transitive on the complex sphere $S^{n-1}_{\CC}$.
 Results on the classification
of the {\it compact} and {\it effective} \footnote{Recall
(see e.g. \cite{MZ} pg. 40) that a transformation group $G$ on a
manifold $M$ is called effective if the only transformation in $G$
that leaves every element of $M$ fixed is the identity in $G$.}
Lie groups transitive on the (real) sphere were obtained
in \cite{Borel} \cite{Samelson1} \cite{Samelson2}. Applications to
control systems were described in \cite{BrockSIAM}. We will
recall in Theorem \ref{3} these results and then will provide further
results and make the necessary connections for
the application of interest here.

\vs We consider the canonical Lie group isomorphism between $U(n)$
and a Lie subgroup of $SO(2n)$. The correspondence between the
matrices $X=R+iY$ in $U(n)$, with $R$ and $Y$ real, and the matrix
$\tilde X \in SO(2n)$ is  given by \be{realificatione} \tilde
X:=\pmatrix{R &- Y \cr Y & R}. \ee The same formula
(\ref{realificatione}) provides the corresponding isomorphism
between the Lie algebra $u(n)$ and a Lie subalgebra of $so(2n)$.
As $X$ acts on $|\psi>:=\psi_R+i\psi_I$ on the complex sphere
$S^{n-1}_{\CC}$, $\tilde X$ acts on the vector $\pmatrix{\psi_R
\cr \psi_I}$ on the real sphere $S^{2n-1}$. Therefore,
transitivity of one action is equivalent to transitivity of the
other. Since $SO(2n)$ is effective on the real sphere $S^{2n -1}$
so is each of its Lie subgroups and in particular the  one
obtained {}from $e^{\cal L}$ via the transformation
(\ref{realificatione}). As for compactness, notice that the
transformation (\ref{realificatione}) preserves compactness.
Moreover, $e^{\cal L}$ is connected and we have the following
facts (see \cite{MZ} pg. 226, we state here this result in a form
suitable to our purposes):

\vs

\bt{2} \cite{MZ} For
every connected
Lie group $G$ which
is transitive on the real sphere, there exists a compact connected
Lie subgroup
$H \subseteq G$ which is also transitive\footnote{Connectedness is
 not explicitly mentioned in the result in \cite{MZ}
but it follows {}from the proof since $H$ is in fact a maximal
compact subgroup of $G$ which is always connected (see \cite{MZ}
pg. 188).}.
\et

\vs

\bt{3} (\cite{Samelson1}, \cite{Samelson2}) The only
compact connected Lie subgroups of $SO(2n)$ that are transitive on
the real sphere of odd dimensions $S^{2n-1}$ are locally
isomorphic to one of the following:
\begin{itemize}
\item[1)]
 $SO(2n)$ itself.
\item[2)] $U(n)$.
\item[3)] $SU(n)$, $n\geq 2$.
\item[4)]
The symplectic group $Sp(\frac{n}{2})$, for $n$ even  and $n>2$.
\footnote{Recall the Lie group of symplectic matrices $Sp(k)$ is
the Lie group of matrices $X$ in $SU(2k)$ satisfying $XJX^T=J$,
with $J$ given by $J=\pmatrix{0 & I_k \cr - I_k & 0}$.}
\item[5)]
The full quaternion-unitary group defined as the group generated by
$Sp(\frac{n}{2})$ and the one dimensional group $\{ K \in U(n) |
K:=e^{i \phi} I_n, \phi \in \RR \}$, $n>2$ and even.
\item[6)] The
covering groups of $SO(7)$ and $SO(9)$ for $n=4$ and $n=8$,
respectively.
\end{itemize}
\et

\vs

Notice that Theorem \ref{3} solves only partially the problem of
determining which subgroups of $SO(2n)$ are transitive on the real
sphere $S^{2n-1}$. In fact,  it only gives a necessary condition for
the Lie algebra to be isomorphic to one of the Lie algebras of the
Lie groups listed in the theorem. It is known that, for example,
the realification (\ref{realificatione}) of the symplectic group
$Sp(\frac{n}{2})$ is  transitive on $S^{2n-1}$, but
nothing can be said {}from the Theorem for Lie groups that are
only locally isomorphic
 (namely have isomorphic Lie algebra)
to $Sp(\frac{n}{2})$, unless further information is supplied. In this
paper we are interested only in the subgroups of $SO(2n)$ that are
isomorphic via (\ref{realificatione}) to a subgroup of $SU(n)$ (or
$U(n)$). We will solve the problem of giving necessary and sufficient
conditions for pure state controllability in terms of
the Lie algebra ${\cal L}$
generated by ${A,B_1,B_2,...,B_m}$ in Theorem \ref{4}. In the following
three Lemmas we use representation theory and structure theory (see
e.g. \cite{Knapp}) to prove three properties of classical Lie groups and
algebras which we will use in the proof of Theorem \ref{4}. We refer to
\cite{Knapp} for the terminology and notions of Lie group theory used
here. We relegate the proofs of the  three lemmas to the
Appendix.

Recalling that, by definition, the covering groups of $SO(7)$ and
$SO(9)$ have Lie algebras isomorphic to  $so(7)$ and $so(9)$
respectively, Lemma \ref{Lem1}  will be used to rule out that
such groups arise, after realification (\ref{realificatione}), as
subgroups of $SU(4)$ (or $U(4)$) and $SU(8)$ (or  $U(8)$).
 \bl{Lem1}\bi \item[(a)] There is no Lie subalgebra of $su(4)$ (or
 $u(4)$)
isomorphic to $so(7)$.  \item[(b)]There is no  Lie subalgebra of
$su(8)$ (or $u(8)$) isomorphic to $so(9)$. \ei \el

 \bl{Lem2}Assume $n$ even. All the subalgebras of $su(n)$ or
$u(n)$ that are isomorphic to $sp(\frac{n}{2})$ are conjugate to
$sp(\frac{n}{2})$ via an element of $U(n)$. \el  \bl{Lem3} Assume $n$ even. Then, the only subalgebra of
$su(n)$ containing $sp(\frac{n}{2})$ (or a Lie algebra isomorphic to
$sp(\frac{n}{2})$)
properly is $su(n)$ itself.
\el

We are now ready to state a necessary and sufficient condition for pure state
controllability in terms of the Lie algebra $\cal L$ generated by
$\{A,B_1,B_2,...,B_m\}$.
\vs
\bt{4} The system is pure state controllable if and only if  ${\cal L}$
is isomorphic (conjugate)  to $sp(\frac{n}{2})$ or to $su(n)$, for $n$
even,  or to $su(n)$, for $n$ odd (with or without the $iI$, where $I$
is the identity matrix).
 \et
\bpr If the system is pure state controllable then $e^{\cal L}$ is
transitive on the complex sphere $S^{n-1}_{\CC}$, therefore its
realification (\ref{realificatione}) is transitive on the real
sphere $S^{2n-1}$. Thus, {}from Theorem \ref{2}, it must contain a
Lie group locally isomorphic to one of the groups listed in
Theorem \ref{3}. As a consequence, the Lie algebra $\cal L$ must
contain a Lie algebra isomorphic to one of the corresponding Lie
algebras. Assume first $n$ odd, then cases 4) 5) and 6) are
excluded. Case 1) is also excluded since $\text{dim} SO(2n) >
\text{dim} U(n)$, when $n \geq 2$ (recall that $SO(2)$ is the
realification of $U(1)$). Therefore $\cal L$ must be either
$su(n)$ or $u(n)$ in this case. If $n=2$ then $su(2)=sp(1)$ so
 cases 3) and 4) and 2) and 5) coincide. If $n$ is
even and $n>2$, then case 1) is excluded as above and cases 2)
through 5) all imply that $sp(\frac{n}{2}) \subseteq {\cal L}$ up
to isomorphism of $sp(\frac{n}{2})$, which {}from Lemma \ref{Lem3}
gives ${\cal L}=sp(\frac{n}{2})$ or ${\cal L}=su(n)$ up to
isomorphism (with or without the identity matrix).
 Case 6)  is
excluded by Lemma \ref{Lem1}. This proves that the only possible
Lie algebras $\cal L$ that correspond to a transitive Lie group
are the ones given in the statement of the Theorem. The converse
follows {}from the well known properties of transitivity of
$SU(n)$ and $Sp(\frac{n}{2})$ as well as of any group conjugate to
them via elements in $U(n)$, and {}from Lemma \ref{Lem2}.
\epr


\section{Equivalent State Controllability}

The notion of equivalent state controllability, although seemingly
weaker, is in fact equivalent to  pure state controllability. In order
to see this, notice that if the system is $ESC$ then for every
pair of states $|\psi_0>$ and $|\psi_1>$ there exists a matrix $X$
in $e^{\cal L}$ and a `phase' $\phi \in \RR$ such that \be{ESC}
X|\psi_0>=e^{i \phi} | \psi_1>. \ee This can be expressed by
saying that there exists an element $Y$ in $e^{i \phi} e^{\cal
L}:= \{Y \in U(n)|Y=e^{i \phi} X, X \in e^{\cal L}, \phi \in
\RR \}$ such that $Y |\psi_0>=|\psi_1>$ and therefore $e^{i \phi}
e^{\cal L}$ is transitive on the complex sphere. Now, if $span\{ i
I_n \} \subseteq {\cal L}$, then $e^{i\phi}e^{\cal L}= e^{\cal L}$
and therefore $e^{\cal L}$ is transitive and the system is $PSC$.
If this is not the case,  then {}{}from Theorem \ref{2}, there
must exist a compact connected Lie group $G \subseteq e^{\cal L}$
such that $e^{i\phi}G$ is transitive. {}From Theorem I' in
\cite{Samelson1}, it follows, writing $e^{i\phi}G$ as $e^{i
\phi}I_n \times G$, that one between the two groups $e^{i\phi}I_n$
and $G$, must be transitive. Therefore $G \subseteq e^{\cal L}$ is
transitive.  In conclusion, we have the following Theorem. \bt{6}
ESC and PSC are equivalent properties for
 quantum mechanical systems (\ref{gensys}).
\et

Theorems \ref{4} and \ref{6} show that a necessary and sufficient
condition to have pure state controllability or equivalent state
controllability is that the Lie algebra ${\cal L}$ is the whole
$su(n)$ or isomorphic to $sp(\frac{n}{2})$ (with or without $iI$).
To check this isomorphism one can apply the structure theory of
Lie algebras to ${\cal L}$. A more practical way to check
equivalent state controllability will be presented in Section 7.
This method only involves elementary matrix manipulations and can
be extended to check density matrix controllability starting from
a fixed given matrix.


\section{Density Matrix Controllability}

 Notice that if $e^{\cal L}=SU(n)$ or $e^{\cal L}=U(n)$ then obviously the system is
 $DMC$. Moreover, in order for the system to be $DMC$,
 the model has to be equivalent state controllable (and therefore
 pure state controllable) as well, because
 transitions
 between pure states represented by matrices of the form
 $|\psi><\psi|$ must be possible. Therefore, to get $DMC$, ${\cal L}$
 must be $su(n)$, or, for $n$ even and
 $n>2$ (see Theorem \ref{4}), it  must be isomorphic (conjugate)
 to $sp(\frac{n}{2})$ (modulo multiples of the identity matrix). The next
 example shows that $Sp(\frac{n}{2})$ is not enough to
 obtain  $DMC$. The example constructs a class of density matrices $D$
 with the property that
 \be{plo76}
 \{WDW^{*} \ |\ W \in Sp(\frac{n}{2}) \} \not=\{ UDU^*|U \in SU(n) \}.
 \ee
 \bex{e1}
 Choose any $n>2$ with $n$ even, and let
$|v>=\left( \begin{array}{c}
          v_{1} \\
          v_{2}
          \end{array} \right) \in \CC^{n}$ and
$|w>=\left( \begin{array}{c}
          -v_{2} \\
          v_{1}
          \end{array} \right) \in \CC^{n}$, with
$v_{1},v_{2}\in \RR^{n/2}$, $||v||=1$. Then
$||w||=1$, $<v|w>=0$, thus, in particular, these two vectors are
independent.
Let
\[
D=\frac{1}{2} \left ( |v><v| + |w><w| \right ).
\]
It is easy to verify that $DJ=JD$ (where $J=\left ( \begin{array}{cc}
    0 & I_{n} \\
    -I_{n} & 0
    \end{array} \right )$). Thus,
if $W\in Sp(\frac{n}{2})$ then we still have that:
\[
\left (WDW^{*} \right) J = J {\overline{ \left (WDW^{*} \right) }}.
\]
Choose any two orthonormal vectors $|v'>, \, |w'> \in \CC^{n}$, such
that:
\[
D' = \frac{1}{2} \left ( |v'><v'| + |w'><w'|  \right ),
\]
satisfies $D'J \neq J\bar{D'}$ (it is easy to see that two such
vectors exist), and let $U\in U(n)$ be any unitary matrix such
that $Uv=v'$ and $Uw=w'$, then
\[
UDU^{*}=D' \neq WDW^{*},
\]
for all $W \in Sp(\frac{n}{2})$. \eex
{}From the above discussion
and example, we can conclude that {\it $DMC$ is equivalent to
$OC$}.

Given a density matrix $D$, it is of interest to give a
criterion on the Lie algebra $\cal L$ for the two orbits \be{orb1}
{\cal O}_{{\cal{L}}}:=\{ W D W^{*} |W \in e^{{\cal L}} \} \ee and
\be{orb2} {\cal O}_{U}:=\{ U D U^{*} |U \in U(n) \} \ee to
coincide. To this aim, notice that since $D$ is Hermitian, $iD$ is
skew-Hermitian so that $iD \in u(n)$, and a matrix commutes with
$iD$ if and only if it commutes with $D$. The {\it centralizer} of
$iD$ is by definition, the Lie subalgebra of $u(n)$ of matrices
that commute with $iD$. Call this subalgebra ${\cal C}_{D}$ and
the corresponding connected Lie subgroup of $U(n)$, $e^{{\cal
C}_{D}}$. Analogously, the centralizer of $iD$ in $\cal L$ is
${\cal C}_{D} \cap {\cal L}$ and we denote by $e^{{\cal C}_{D}
\cap {\cal L}}$ the corresponding subgroup of $U(n)$ (which is
also a subgroup of $e^{{\cal L}}$).

For a given density matrix $D$, it is sufficient to calculate  the
dimensions of ${\cal L}$, ${\cal C}_{D}$ and ${\cal C}_{D}\cap
{\cal L}$ to verify the equality of the two orbits ${\cal
O}_{{\cal{L}}}$ and ${\cal O}_{U}$ defined in (\ref{orb1})
(\ref{orb2}). We have the following  result.
 \bt{pp1} Let $D$ be a
given density matrix, then ${\cal O}_{{\cal{L}}}={\cal O}_{U}$ if
and only if \be{u3} \dim u(n)- \dim {\cal C}_{D}= \dim  {\cal L} -
\dim ({\cal L} \cap {\cal C}_{D}). \ee \et \bpr We have the
following isomorphisms between the two coset spaces
$U(n)/{e^{{\cal C}_{D}}}$  and  $e^{{\cal L}}/{e^{{\cal C}_{D}\cap
{\cal L}}}$ and the two manifolds  ${\cal O}_{U}$ and ${\cal
O}_{{\cal{L}}}$, respectively: \be{u1}
{U(n)}/{e^{{\cal C}_{D}}} \simeq \{ UDU^{*} \ | \ U\in U(n) \},
\ee \be{u2} {e^{\cal L}}/{e^{{{\cal C}_{D}}\cap {\cal L}}} \simeq
\{ WDW^{*} \ | \ W\in e^{\cal L} \}, \ee where $\simeq$ means
isomorphic. Therefore if the two orbits coincide, we must have
that  the two coset spaces must coincide as well. So, in
particular, their dimensions have to be equal which gives
(\ref{u3}).

Conversely assume that (\ref{u3}) is verified. Then the dimensions
of the two coset spaces on the left hand sides of (\ref{u1}) and
(\ref{u2}) are the same and so are the dimensions of the manifolds
on the right hand side namely ${\cal O}_{U}$ and ${\cal
O}_{{\cal{L}}}$. Notice also that these two manifolds are
connected since both $U(n)$ and $e^{{\cal L}}$ are connected.
Since $e^{{\cal C}_{D}}$ is closed in $U(n)$ and therefore
compact, {}from Proposition 4.4 (b) in \cite{Helgason} we have
that $e^{\cal L}/e^{{\cal L}\cap {\cal C}_{D}}$ is closed in
$U(n)/e^{{\cal C}_{D}}$. On the other hand, since the two coset
spaces have the same dimensions, $e^{\cal L}/e^{{\cal L}\cap {\cal
C}_{D}}$ is  open in $U(n)/e^{{\cal C}_{D}}$. By connectedness, we
deduce that the two coset spaces must coincide, and therefore
the two orbits coincide as well. \epr

Special cases of the above Theorem, are density matrices
representing pure states or completely random states. In the first
case, the density matrix $D$ as the form, $D=|\psi><\psi|$ and, in
an appropriate basis, it can be written as a diagonal matrix with the
$(1,1)$ entry equal to one and  all the remaining entries equal to
zero. The analysis
in Section 4 shows that the only Lie algebras ${\cal L}$
satisfying condition (\ref{u3}) are $su(n)$  or,
for $n$ even,
isomorphic to $sp(\frac{n}{2})$ (with or without $iI$). For completely
random states, the density matrix $D$ is a real scalar matrix with
trace equal to one, and therefore its centralizer in ${\cal L}$,
${\cal L}\cap {\cal C}_{D}$, is all of ${\cal L}$, for every
subalgebra ${\cal L}$. Thus the condition (\ref{u3}) holds with
$\dim {\cal L} - \dim { {\cal L} \cap {\cal C}_{D}} =0$ for every
${\cal L}$. The interpretation, {}from a physics point of view, is
the obvious fact that a completely random ensemble of quantum
systems remains completely random after any evolution. The paper
\cite{Schirmer3} contains a complete classification
of density matrices as well as additional results on density matrix
 controllability.


\section{Test of Controllability}

As we have shown in the previous sections, the two notions of
operator-controllability (in the special unitary case)
and density-matrix-controllability are
equivalent and they are the strongest among the controllability
notions we have defined. On the other hand, pure state-controllability
and equivalent-state-controllability are equivalent. These facts
are summarized in the following diagram:

\[
\text{ DMC } \ \ \Leftrightarrow \ \ \text{ OC } \ \ \Rightarrow \ \ \text{
PSC } \ \ \Leftrightarrow \ \ \text{ ESC }.
\]
 {}From a practical point of view, it is of great interest to
give criteria on the Lie algebra $\cal L$ to ensure that the
corresponding group is transitive on the complex sphere. In this
case the system is pure state controllable. As we have seen {}from the
analysis in Section 4, the Lie algebra $\cal L$ has to be to
$su(n)$ or $u(n)$ or, for $n$ even,  conjugate and therefore
isomorphic to $sp(\frac{n}{2})$ (modulo multiples of the identity).
To check this isomorphism, one
can apply the Cartan theory of classification of semisimple Lie
algebras \cite{Helgason}. A simpler and self contained
test can be derived {}from
Theorem \ref{pp1}. To this purpose, notice that pure state
controllability is the same as equivalent state controllability
(see Theorem \ref{6}) and this can be easily seen to be equivalent
to the possibility of steering the matrix \be{matricepura} D=
\text{diag}(1,0,0,...,0) \ee to any unitarily  equivalent matrix.
The centralizer ${\cal C}_{D}$ of the matrix $iD$ in
(\ref{matricepura}) in $u(n)$,  is given by the set of matrices of
the form \be{formmm} M:=\pmatrix{ia & 0\cr 0 & H}, \ee with $a$
any real and $H$ a matrix in $u(n-1)$. The dimension of ${\cal
C}_{D}$ is $(n-1)^{2}+1$ and therefore the number on the right
hand side of (\ref{u3}) is $n^{2}-((n-1)^{2}+1)=2n-2$. In
conclusion as a consequence of Theorems \ref{pp1} and \ref{6} we
have the following easily verifiable criterion for pure state
controllability.

\bt{crit} With the above notations and definitions, the system
(\ref{jasbd}) is pure state controllable if and only if the Lie algebra
$\cal L$ generated by $\{A, B_{1}, B_{2},\ldots,B_{m}\}$ satisfies
\be{hh} \dim{\cal L}- \dim({\cal L} \cap {\cal C}_{D})=2n-2. \ee
\et

We remark here that similar criteria can be given for
different density matrices according to Theorem \ref{pp1}.

\bex{whatever} Assume that the Lie algebra $\cal L$ is given by
the matrices of the form \be{forrrr} F:=\pmatrix{L+Z & T+C \cr
-\bar T + \bar C &-L+Z^T}, \ee with $L$ diagonal and purely
imaginary, $T$ diagonal, and $Z, \, C$ having zeros on the main
diagonal, all of them $2 \times 2$ matrices. This Lie algebra is
in fact conjugate to $sp(2)$. Verifying this fact directly can be
cumbersome. However to prove that the associated system is pure state
controllable, one can verify that the Lie subalgebra of matrices
of  $\cal L$ that  have the form (\ref{formmm}), namely ${\cal L}
\cap {\cal C}_D$,   has dimension $4$. Since the dimension of
$\cal L$ is $10$, we have (recall $n=4$) \be{finallopolkjilo} \dim
{\cal L} - \dim {\cal L} \cap {\cal C}_D=6=2n-2. \ee Therefore the
criterion of Theorem \ref{crit} is verified.

 \eex

\section{State transfer in arbitrary time}

In this section, we study the transfer of  state of
(\ref{gensys})
and (\ref{jasbd}) in arbitrary small time. For the results that will follow,
we can  assume that $X$ varies on a general compact transformation (matrix)
Lie group $G$ (with
corresponding subalgebra $\cal G$) while
$|\psi>$ varies on the corresponding homogeneous
space $M$.
First, we define the set of states reachable in arbitrary time for
system (\ref{jasbd}). We denote it by ${\cal A}\subseteq G$; we have:
\be{plo}
{\cal A}:=\cap_{t>0} R(t),
\ee
where $R(t)$ is the set of states
reachable {}{}from the identity at time
$t$ for system (\ref{jasbd}).
As it has been  shown with  a number of examples \cite{ioSCL}
\cite{SUSS}, even though the set of states reachable {}from the
identity for (\ref{jasbd}) is  the whole group $G$ and the
magnitude of the controls is unconstrained, it is possible that
not all the states in $G$
 can be obtained in arbitrary small time. In fact, conditions can be
given on $A$, $B_1$,...,$B_m$ for ${\cal A}$
to be empty \cite{MikoCDC}. However, it may well be that even if
${\cal A}$ is a proper subset of  $G$, ${\cal A}$ is still
transitive on $M$. This fact is easily seen to be necessary and sufficient
for state transfer in arbitrary small
time between two states in $M$.

In this section, we investigate the relation between
controllability in arbitrary time on $M$ and $G$ and then apply this
result to quantum control systems.

\vs

Instead
of working with $\cal A$ in (\ref{plo}), it is
more convenient to work with its `regularized version'
\be{agy}
{\cal A}_{reg}:=\cap_{t>0} \bar R(t),
\ee
where $\bar R(t)$ is the closure of $R(t)$. The set ${\cal A}_{reg}$ has
more structure because it is a compact connected Lie subgroup of $G$
\footnote{It is proved in \cite{MikoCDC} that Small Time Local Controllability
of the identity of the group $G$ implies ${\cal A}={\cal A}_{reg}$.}
\vs
\bt{agg1} If ${\cal A}$ is not empty,
${\cal A}_{reg}$ is a compact connected Lie subgroup of $G$.
\et
\bpr
First notice that ${\cal A}_{reg}$ is a semigroup. In fact, assume $X_1$
and $X_2$ are in $\bar R(\frac{t}{2})$ for every $t$. Then there exist
two sequences of elements $\{X_{1k}\}$ and $\{X_{2k}\}$ in $R(\frac{t}{2})$
converging to $X_1$ and $X_2$, respectively. The elements of the
sequence  $\{ X_{1k}X_{2k} \}$ are all in $R(t)$ and by continuity
$\lim_{k \rightarrow \infty}X_{1k}X_{2k}=X_1X_2,$
so that $X_1X_2 \in \bar R(t)$. Since $t$ is arbitrary, this proves
$X_1X_2 \in {\cal A}_{reg}$. Consider now an element $X \in {\cal
A}_{reg}$. $X^n$ is also in ${\cal A}_{reg}$, for every $n$, and by
compactness, the sequence of $X^n$'s has a converging subsequence
$\{X^{n(k)}\}$. The sequence $X^{n(k+1)-n(k)-1}$ converges as $k$ tends to
infinity to $X^{-1}$ and therefore $X^{-1} \in {\cal A}_{reg}$
(cfr. \cite{SUSS}). Since
${\cal A}_{reg}$ is a closed subgroup of the Lie group $G$, it is itself
a Lie subgroup (\cite{warner} pg. 110). To prove connectedness notice
that, if $\cal A$ is not empty, $t_1 < t_2$ implies
$\bar R(t_1) \subseteq R(t_2)$ (see \cite{MikoCDC} Theorem
3.2). Therefore ${\cal A}_{reg}$ is the intersection of a decreasing
sequence of compact and connected sets $\bar R(t)$ (connectedness of
these sets was proven in \cite{SUSS}); thus ${\cal{A}}_{reg}$ is itself
compact and
connected (\cite{kuratowskii} pg. 212).
\epr

If $\cal A$ is not empty, then
 ${\cal A}_{reg}$ is a connected Lie group and  we can consider its
associated Lie algebra ${\cal L}_A$. Consider $\cal B$ the Lie algebra
generated by $B_1,...,B_m$ in (\ref{jasbd}). We have.

\bt{hgf}
${\cal B} \subseteq {\cal L}_A$
\et
\bpr
It follows immediately {}from the fact (see Theorem 3.3 in \cite{MikoCDC})
 that the connected subgroup corresponding to $\cal B$ is a subgroup of
${\cal A}_{reg}$.
\epr

In the following theorem we will call transitive a subalgebra of $\cal
G$ whose corresponding Lie subgroup of $G$ is transitive on $M$. We have

\bt{final}
Assume $\cal B$ is not a subalgebra of any transitive proper subalgebra of
$\cal G$. If the system is state controllable in arbitrary time,
then ${\cal A}_{reg}=G$, in particular for every $t$ $\bar R(t)=G$.
\et
\bpr
Assume that the system is state controllable in
arbitrary time. This means that the set
$\cal A$ is transitive on $M$
and since ${\cal A} \subseteq {\cal A}_{reg}$, so is
${\cal A}_{reg}$. It follows {}from the assumptions on $\cal B$ and the
fact that ${\cal B} \subseteq {\cal L}_A$ that ${\cal {A}}_{reg}$ has to
be equal to $G$.
\epr

\vs

In   \cite{CDC},  \cite{ACC}, \cite{RamaNHP} and \cite{BKS} the system of two
 interacting spin $\frac{1}{2}$ interacting particles in a driving
electro-magnetic field was considered. The system has the form
(\ref{gensys}), with $m=3$, where the matrices $A,$ $B_1$,$B_2$,
$B_3$,  are appropriate matrices in $su(4)$ and the solution
$|\psi>$ varies on the sphere $S_{\CC}^3$ while
 $X$ in (\ref{jasbd}) varies in $SU(4)$.
The matrix $A$ models the interaction between the two particles which
can assume different forms (e.g. isotropic, dipolar) while $B_{1}$,
$B_{2}$ and $B_{3}$ model the interaction between particles and
the external field.
 The controls
$u_1$, $u_2$, $u_3$ are component of a driving electro-magnetic field in
the $x$, $y$ and $z$
direction, respectively. In typical experimental set ups, the
$z$-component of the field is held constant.
  This system is of interest because it is used to perform two quantum
bit logic operations in quantum computing \cite{divi}. For a given
matrix $U \in SU(4)$,  define $T_{U}:=\inf \{ t \geq 0 \ | \ U \in R(t)
\}$. It follows from the results in \cite{BKS} that $T_U$ is not zero
for every $U \in SU(4)$ and, as a consequence, not every matrix in
$SU(4)$ can be reached  in arbitrary small time. The Lie algebra $\cal
B$ generated by $B_1$, $B_2$, and $B_3$ is in this case conjugate to
$so(4)$ which satisfies the same property stated for
$sp(2)$ in Lemma
\ref{Lem3}. Moreover, since
$so(4) \not= sp(2)$, it follows  from Theorem 4 that $\cal B$ is not a
transitive subalgebra.  Therefore, the system satisfies the conditions of
Theorem \ref{final}, and it follows that the system is not state
controllable in arbitrary time. In fact, if this was the case,
then ${\cal A}_{reg}=SU(4)$ or equivalently
for every time $t$, $\bar R(t)=SU(4)$. On the other
hand \be{implicazione} \bar R (t)=SU(4) \ \ \Rightarrow  \ \
T_{U}=0 \text{ for all } U, \ee which is false, as seen before. To
prove (\ref{implicazione}), we argue by contradiction. Assume
that, for a given $U\in SU(4)$,  $T_{U}>0$. Since $SU(4)=
\text{int}\left( R^{-1}(T_{U}/2)U\right)$ and $SU(4)=\bar
R(T_{U}/4)$, there exists an open set $N$ such that:
\[
N\subseteq \text{int}\left( R^{-1}(T_{U}/2)U\right) \cap
R(T_{U}/4).
\]
Choose any matrix $\tilde{U}\in N$, then {}from $I$ we reach
$\tilde{U}$ in time $T_{U}/4$, and {}from $\tilde{U}$ we reach $U$
in time $T_{U}/2$, thus $U\in R(\frac{3T_{U}}{4})$, which
contradicts the minimality of $T_{U}$.

\section{Conclusions}

For bilinear quantum mechanical systems in the multilevel approximation a
number of concepts concerning controllability can be considered.
One can ask whether it is possible to drive the evolution
operator or the state to any desired configuration. One
typically represents the state with a vector with norm 1 or using
the density matrix formalism. Connections between different
notions of controllability have been established in this paper,
where we have shown that the possibility of driving a pure state
between two arbitrary configurations is in general a weaker
property than the controllability of the evolution operator. All
the controllability properties of a given quantum system can
studied by studying the Lie algebra generated by the matrices
$\{ A, \, B_{1}, \ldots, B_{m}\}$ of the system (\ref{gensys}).
This Lie algebra has to be the
full Lie algebra $su(n)$ (or $u(n)$) for controllability of the
operator while for controllability of the state it can be
conjugate and therefore isomorphic to the Lie algebra of
symplectic matrices of dimension $n$ modulo a phase factor.
We have also given a
practical test to check this isomorphism. This test can be
extended for density matrices of rank different {}from one and
only requires elementary algebraic manipulations involving the
centralizer of the given density matrix.

The paper also contains some results on the relation between
controllability in arbitrary small time for a system on a Lie
transformation group and for the corresponding system on the
associated homogeneous space. The application of these results  to
the systems of two interacting spin $\frac{1}{2}$ particles in an
electro-magnetic field shows the negative result that it is not
possible for this system to obtain a state transfer between two
points even though the control can be taken of arbitrary large
magnitude.

\section*{Appendix: Proofs of Lemmas 4.1, 4.2 and 4.3}

\subsection*{Proof of Lemma 4.1}
First
notice that neither $so(7)$ nor $so(9)$ have an element which
commutes with all the algebra, therefore if there exists a
subalgebra of $u(4)$ (resp. $u(8)$) isomorphic to $so(7)$ (resp.
$so(9)$), it must be also a subalgebra of $su(4)$ (resp. $su(8)$).

Statement (a) of the Lemma can be checked by calculating the dimensions
of $su(4)$ and $so(7)$. We have $dim(su(4))=15<dim(so(7))=21$. As
for statement (b), assume there exists a subalgebra of
$su(8)$, call it $\cal F$,
 isomorphic to $so(9)$, namely a
(faithful) representation of $so(9)$. Assume first this representation
to be irreducible. Then  there is an highest associated
weight by the fundamental theorem of representation theory
(see e.g. \cite{Knapp}, Theorem 4.28). The basic
weights are given by
$w_1=(0,0,0,0)$,
$w_2=(\frac{1}{2},\frac{1}{2},\frac{1}{2}, \frac{1}{2})$,
$w_3=(1,0,0,0)$, $w_4=(1,1,0,0)$, $w_5=(1,1,1,0)$,
$w_6=(1,1,1,1)$. $w_1$ corresponds to the trivial representation which
is obviously not faithful. For each one of the others the underlying
vector space $V$ on which the representation acts has dimension that can
be calculated using Weyl formula (see e.g. \cite{SW} pg 332). This
calculation gives the following values for $dim(V)$, for the cases
$w_2$, $w_3$, $w_4$, $w_5$, $w_6$, respectively, $16$, $9$, $36$,
$84$, $252$. In any case, the dimension is bigger than $8$ which is the
maximum allowed by the fact that $\cal F$ is a subalgebra of
$su(8)$. All the other irreducible transformations can be calculated as
tensor products of the basic representations (\cite{Knapp} Pr.11
pg. 111, \cite{SW} Corollary 15.18 pg, 330) and therefore the dimension
of $V$ in this case is the product of the dimensions of the basic
representations and therefore $>8$. If the representation is not
irreducible then it is the direct sum of irreducible transformations
(\cite{Knapp} pg. 15 Corollary 1.7) and therefore the vector space $V$
has dimension which is the sum  of sum above given. In this case, the
only possibility to have $dim(V)\leq 8$ is that $V=\oplus_{j=1}^r V_j$ and
the representation acts as the trivial representation on any $V_j$,
which makes it not faithful. \qed

\subsection*{Proof of Lemma 4.2}

Consider a
subalgebra ${\cal F}\subseteq u(n)$ isomorphic to
$sp(\frac{n}{2})$. It follows immediately {}from the fact that
$\cal F$ is semisimple that $iI \notin {\cal F}$ and therefore
${\cal F} \subseteq su(n)$. Thus,  $\cal F$ is a faithful
representation of $sp(\frac{n}{2})$. Assume first that this
representation is irreducible. Consider the parametrizations of
the finite dimensional representations of $sp(\frac{n}{2})$ given
by the theorem of the highest weight (see e.g. \cite{Knapp}
Theorem 4.28). The $n-dimensional$ basic weight vectors are
$w_1=(0,0,...,0)$,
$w_2=(1,0,...,0)$,$w_3=(1,1,0,...,0)$,...,$w_{\frac{n}{2}+1}=(1,1,1,...,1)$.
$w_1$ gives the trivial representation which is not faithful;  the
representation corresponding to $w_2$ acts on a vector space $V$
of dimension $n$. All the other representations act on vector
spaces $V$ of dimension $> n$. The same is true for the other
irreducible representations whose weights are sums of some $w_j$,
$j=1,...,\frac{n}{2}+1$. As for reducible representations they are
sums of the irreducible ones and therefore the dimension of the
underlying vector space $V$ is $> n$ except for the sum of a
number $l \leq n$ of trivial representations which is a (higher
dimensional) trivial representation and clearly not faithful.
Therefore the only possible representations of dimensions $n$ are
all equivalent to each other and in particular they are equivalent
to the basic representation of $sp(\frac{n}{2})$. In conclusion
there exists a nonsingular matrix $E$ such that \be{equivalenza}
{\cal F}=E sp(\frac{n}{2})E^{-1}. \ee Notice that $E$ is defined
up to a multiplicative constant. It remains to show that $E$ can
be chosen in $U(n)$. The connected Lie subgroup of $SU(n)$ with
Lie algebra ${\cal F}$ is a unitary representation  of
$Sp(\frac{n}{2})$ that assigns to an element $g\in
Sp(\frac{n}{2})$ an element $\Phi(g)$ and, {}from
(\ref{equivalenza}), we have \be{lplplplp} E=\Phi^*(g)Eg, \ee
{}from which it follows \be{plpoiuytrewq}
EE^*=\Phi^*(g)EE^*\Phi(g), \ee and \be{gfddfg}
\Phi(g)EE^*=EE^*\Phi(g). \ee The matrix $EE^*$ commutes with all
the elements of a unitary irreducible representation and therefore
{}from Schur's Lemma (see e.g. \cite{Knapp} Proposition 1.5) it
must be a scalar matrix $\alpha I$, with $\alpha$ real $>0$. Thus,
scaling $E$ by a factor $\sqrt{\alpha}$ we can make $E$ unitary.
\qed

\subsection*{Proof of Lemma 4.3}

It follows from the results in \cite{Dynkin} that the complexification
of $sp(\frac{n}{2})$ is a maximal subalgebra in the complexification of
$su(n)$. Now, if there exists a proper subalgebra $\cal F$ of $su(n)$
properly containing $sp(\frac{n}{2})$, then its complexification will be
a proper subalgebra of the complexification of $su(n)$ properly
containing the complexification of $sp(\frac{n}{2})$ (cfr. \cite{SW}
Section 9.3) which contradicts the maximality
of $sp(\frac{n}{2})$.  \qed

\end{document}